\documentclass[11pt,a4paper]{article}
\usepackage[english]{babel}
\usepackage[mathscr]{eucal}
\usepackage{amsmath,amsbsy,amsthm,amsfonts,amssymb,makeidx,enumerate,bm,wasysym}
\usepackage{courier,vmargin,graphicx,wrapfig,float}
\usepackage{caption,subcaption}
\usepackage{epstopdf}

\numberwithin{equation}{section}

\setmarginsrb{2.2 cm}{1 cm}
             {2.2 cm}{1.5 cm}
             {1.2 cm}{1. cm}
             {1.2 cm}{0.8 cm}

\def\al{\alpha}
\def\be{\beta}
\def\ga{\gamma}
\def\om{\omega}

\def\Z{\mathbb{Z}}
\def\R{\mathbb{R}}
\def\ff{\varphi}
\def\XX{\mathcal{X}}
\def\geqs{\geqslant}
\def\leqs{\leqslant}
\def\dd{\displaystyle}
\def\To{\mathbf{T}}
\def\ellm{\lambda}
\def\ellp{\Lambda}
\def\HH{\mathcal{H}}

\def\yr{y}
\def\Rs{\mbox{R}}
\def\Ric{\mbox{Ric}}
\def\Riem{\mbox{Riem}}
\def\TT{\mathfrak{T}}
\def\de{\partial}
\def\ze{\mbox{{\small $(0)$}}}
\def\xf{\varpi}
\def\AAt{\mathcal{A}_{\tau}}
\def\aa{\alpha}
\def\ac{\mathrm{a}}
\def\gt{\mbox{g}_{\earth}}
\def\Te{\mbox{T}}
\def\EE{\mathcal{E}}
\def\EFg{\mathfrak{E}_\xi}
\def\EFf{\mathfrak{E}_f}
\def\me{\mbox{m}}
\def\cm{\mbox{cm}}
\def\gr{\mbox{gr}}
\def\ye{\mbox{y}}
\def\se{\mbox{s}}
\def\JJ{\mathbf{J}}
\def\II{\mathbf{1}}
\def\PP{\mathbf{P}}
\def\QQ{\mathbf{Q}}
\def\zer{\mathbf{0}}
\def\ka{\mu}

\title{\textbf{Some remarks on a new exotic spacetime\vspace{-0.05cm}\\ for time travel by free fall}}

\author{
Davide Fermi \\
Dipartimento di Matematica, Universit\`a di Milano\\
Via C. Saldini 50, I-20133 Milano, Italy\\
and Istituto Nazionale di Fisica Nucleare, Sezione di Milano, Italy \vspace{0.2cm}\\
e--mail: davide.fermi@unimi.it
}

\date{}

\begin{document}
\maketitle

\begin{abstract} 
This work is essentially a review of a new spacetime model with closed causal curves, recently presented in another paper (Class.\! Quantum Grav.\! \textbf{35}(16) (2018), 165003).
The spacetime at issue is topologically trivial, free of curvature singularities, and even time and space orientable. Besides summarizing previous results on causal geodesics, tidal accelerations and violations of the energy conditions, here redshift/blue\-shift effects and the Hawking-Ellis classification of the stress-energy tensor are examined.
\end{abstract}
\vspace{0.2cm} \noindent
\textbf{Keywords:} closed causal curves, time machines, energy conditions, Hawking-Ellis classification.
\vspace{0.15cm}\\
\textbf{MSC\! 2010}: 83C99, 83-06\,.
\vspace{0.15cm}\\
\textbf{PACS\! 2010}: 01.30.Cc, 04.20.Cv, 04.20.Gz, 04.90.+e\,.
\vspace{0.2cm}

\section{Introduction}
In spite of being inherently locally causal, the classical theory of general relativity is well-known to encompass violations of the most intuitive notions of chronological order and causality. Such violations are generically ascribable to the fact that ``faster than light'' motions and ``time travels'' (namely, closed causal curves) are not ruled out by first principles.
Several spacetime models sharing the above exotic features are known nowadays; see \cite{KraB,Lobo,LoboB,ThoRev} for comprehensive reviews. In \cite{CQG} a four-fold classification of the existing literature was suggested:\\
\textsl{First class}. Exact solutions of the Einstein's equations, where closed timelike curves (CTCs) are produced by strong angular momenta or naked singularities. These include the G\"odel metric \cite{God}, the Van-Stockum dust \cite{Tip,VanS}, the Taub-NUT space \cite{Mis,Mis2,NUT,Taub,ThoM}, the Kerr black hole \cite{Hawk,Kerr,Tom35}, the radiation models of Ahmed \emph{et al.} \cite{Ahm2,Ahm1}, as well as certain cosmic strings models \cite{Des,Gott,Grant,sing}.\\
\textsl{Second class}. Specifically designed Lorentzian geometries, describing CTCs at the price of violating the usual energy conditions. Examples of such geometries are those analysed by Ori and Soen \cite{Scher,Ori1,Ori3,Ori2,Ori4}, and by Tippett and Tsang \cite{tardis}.\\ 
\textsl{Third class}.\! \emph{Ad hoc} spacetimes for ``faster than light'' motions, violating the standard energy conditions. These include the Alcubierre warp drive \cite{Alcu}, the Krasnikov-Everett-Roman tube \cite{Eve1,Eve2,Kras} and the Ellis-Morris-Thorne wormhole \cite{Ell,MT1,MT2}.\\
\textsl{Fourth class}. A single model by Ori \cite{Ori2007} which presents CTCs, still fulfilling all the classical energy conditions. The corresponding metric is obtained evolving via Einstein's equations some suitable Cauchy data, corresponding to dust-type matter content; unluckily, this characterization of the metric is not explicit enough to rule out the appearance of pathologies (e.g., black holes).\vspace{0.05cm}\\
Many problematic aspects are known to affect spacetimes with CTCs. For example, the well-posedness plus existence and uniqueness of solutions for various kinds of Cauchy problems is a rather delicate issue \cite{Bache,Eche,Frie,MNT,NovB}. Besides, divergences often arise in semiclassical and quantized field theories living on the background of spacetimes with CTCs \cite{CPC,KDiv,Viss}; arguments of this type led Hawking to formulate the renowned ``chronology protection conjecture'' \cite{CPC}.\\
This work reviews a spacetime of the second class, recently proposed in \cite{CQG}. In the spirit of Alcubierre's and Krasnikov's interpolation strategies \cite{Alcu,Kras}, the said spacetime is intentionally designed as a smooth gluing of an outer Minkowskian region, and an inner flat region with CTCs; the topology is trivial and no curvature singularity occur.
The geometry of the model is established in Section \ref{SecGeo}; natural time and space orientations are described in Section \ref{SecOr}. Some manifest symmetries of the model are pointed out in Section \ref{SecSym} and then employed in Section \ref{quadSec} for reducing to quadratures the equations of motion for a specific class of causal geodesics. Section \ref{ffSec} discusses the existence of causal geodesics connecting a point in the outer Minkowskian region to an event in its causal past. Tidal accelerations experienced by an observer moving along a timelike geodesic of the latter type are discussed in Section \ref{secTid}, while frequency shifts of light signals propagating along null geodesics are analysed in Section \ref{RedSec}. The matter content of the spacetime is derived in Section \ref{mattSec}; there it is shown that the classical energy conditions are violated and the Hawking-Ellis classification of the stress-energy tensor is determined.
Finally, Section \ref{secfut} mentions some possible developments, to be analysed in future works.

\section{Postulating the geometry}\label{SecGeo}
Assume \emph{ab initio} that the geometry of the spacetime is described by the Lorentzian manifold
\begin{equation}
\TT := (\R^4,g)\,, \label{TTdef}
\end{equation}
where $g$ is the metric defined by the line element
\begin{equation}
ds^2 = -\,\big[(1\!-\!\XX)\,dt + \XX a R\,d\ff\big]^2\! + \big[(1\!-\!\XX)\,\rho\,d\ff - \XX \,b\,dt\big]^2\!
+ d\rho^2\! + dz^2\, . \label{dsXidef}
\end{equation}
The above expression for $ds^2$ is characterized by the following building blocks:
\vspace{-0.1cm}
\begin{enumerate}[(i)]
\item a set of cylindrical-type coordinates $(t,\ff,\rho,z) \in \R \times \R/(2\pi \Z) \times (0,+\infty)\times \R$ on $\R^4$ (notice that units are fixed so that the speed of light is $c = 1$);
\vspace{-0.1cm}
\item a pair of dimensionless parameters $a,b \in (0,+\infty)$, playing the role of scale factors (cf. Eq. \eqref{ds1def}, below);
\vspace{-0.1cm}
\item two concentric tori $\To_{\ellm},\To_{\ellp}$ in $\R^3$ (for fixed $t \in \R$), with common major radius $R$ and minor radii $\ellm,\ellp$ such that $\ellm < \ellp < R$, namely, (see Fig. 1)
\begin{equation}
\To_{\ell} := \big\{ \sqrt{(\rho-R)^2 + z^2} = \ell\,\big\} \qquad\;
\big(\ell = \ellm,\ellp\big)\,; \label{Todef} \vspace{-0.1cm}
\end{equation}
\item an at least twice continuously differentiable function $\XX \!\equiv\! \XX(\rho,z)$, such that
\begin{equation}
\XX = \left\{\!\begin{array}{ll}
1 & \mbox{inside $\To_{\ellm}$}\,,\\
0 & \mbox{outside $\To_{\ellp}$}\,.
\end{array}	\right.
\end{equation}
For later convenience, we fix
\begin{equation}
\XX(\rho,z) := \HH\big(\sqrt{(\rho/R-1)^2+(z/R)^2}\,\big) \,, \label{Xidef}
\end{equation}
where $\HH$ is a shape function on $[0,+\infty)$ of class $C^k$ for some $k \in \{2,3,...,\infty\}$, such that $\HH(\yr) = 1$ for $0 \!\leqs\! \yr \!\leqs\! \ellm/\!R$, $\HH(\yr) = 0$ for $\yr \!\geqs\! \ellp/\!R$ and $\HH'(\yr) \!<\! 0$ for $\ellm/\!R \!<\! \yr \!<\! \ellp/\!R$ (piecewise polynomial choices of $\HH$ fulfilling these requirements are described in \cite[App. A]{CQG}; see also the forthcoming Eq. \eqref{cho1} and Fig. 2).
\newpage
\noindent
\end{enumerate}
\begin{figure}[t!]\label{fig:ToriHk}
\vspace{-0.7cm}
    \centering
        \begin{subfigure}[b]{0.475\textwidth}\label{fig:Tori}
                \includegraphics[width=\textwidth]{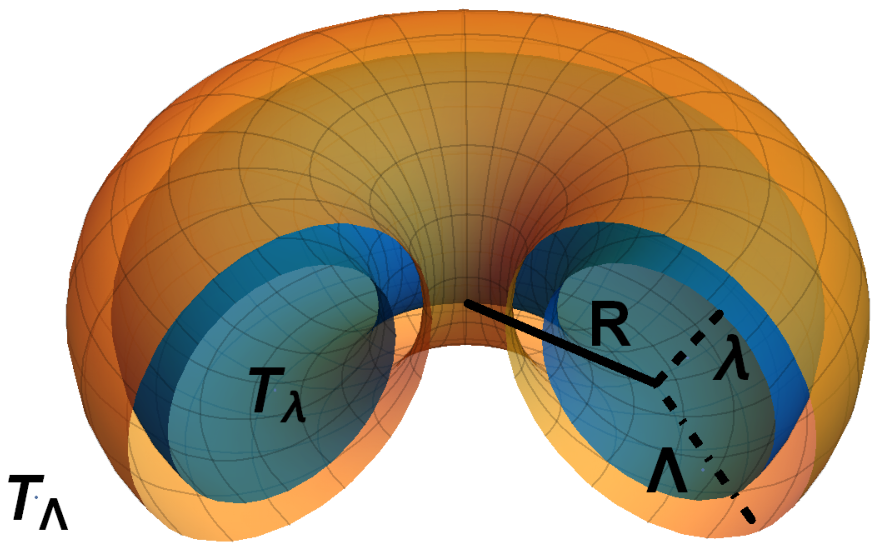}\vspace{-1.5cm}
                \caption*{\textsc{Figure 1.} Representation (for fixed $t$) of the tori $\To_{\ellm}$ (in blue)
                and $\To_{\ellp}$ (in orange).}
        \end{subfigure}
        \hspace{0.4cm}
        \begin{subfigure}[b]{0.475\textwidth}
                \includegraphics[width=\textwidth]{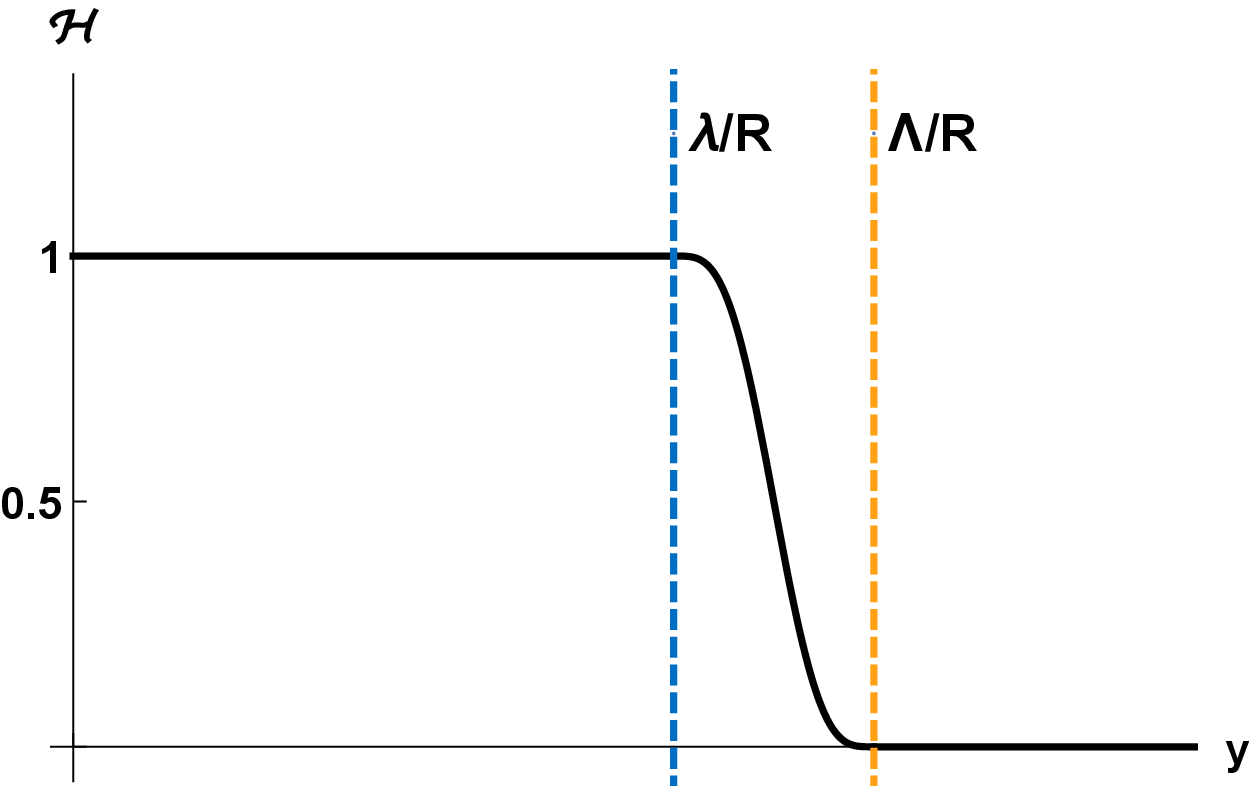}\label{fig:Hk}
                \caption*{\textsc{Figure 2.} Plot of an admissible shape function $\HH$ (cf. Eq. \eqref{cho1}).}
        \end{subfigure}
\vspace{-0.15cm}
\end{figure}
Paying due attention to the usual coordinate singularity at $\rho = 0$, it can be checked that the line element \eqref{dsXidef} does indeed define a non-degenerate, symmetric bilinear form of constant signature $(3,1)$, i.e., a Lorentzian metric $g$ on the whole spacetime manifold $\R^4$ (see \cite{CQG} for more details). Furthermore, on account of the presumed regularity features of $\XX$, the said metric $g$ is granted by construction to be of class $C^k$, with $k\!\geqs\! 2$; this suffices to infer that the corresponding Riemann curvature tensor $\Riem_g$ is of class $C^{k-2}$ (whence, at least continuous) and, in particular, free of singularities, alike all the associated curvature invariants.
\\
Next, notice that outside the larger torus $\To_{\ellp}$ (where $\XX \equiv 0$) the line element $ds^2$ of Eq. \eqref{dsXidef} reproduces the usual Minkowskian analogue
\begin{equation}
ds_0^2 := -\,dt^2\! + \rho^2\,d\ff^2\! + d\rho^2\! + dz^2\, . \label{ds0def}
\end{equation}
On the other hand, inside the smaller torus $\To_{\ellm}$ (where $\XX \equiv 1$), $ds^2$ matches the flat line element
\begin{equation}
ds_1^2 := -\,a^2 R^2 d\ff^2\! + b^2\,dt^2\! + d\rho^2\! + dz^2\, . \label{ds1def}
\end{equation}
This shows that inside $\To_{\ellm}$ the periodic variable $\ff \!\in\! \R/(2\pi \Z)$ is a time coordinate, meanwhile $t \in \R$ is a space coordinate: the opposite of what happens in the Minkowskian region outside $\To_{\ellp}$. In consequence of this, within $\To_{\ellm}$ the curves naturally parametrized by $\ff$ (with $t,\rho,z$ fixed) are indeed geodesic CTCs.
\\
As a matter of fact, the line element \eqref{dsXidef} is designed on purpose as a sort of interpolation (parametrized by the the shape function $\XX$) between the two flat line elements \eqref{ds0def} \eqref{ds1def}; this is achieved by means of a strategy similar to those originally devised by Alcubierre \cite{Alcu} and Krasnikov \cite{Kras}, and later successfully adopted by other authors for the construction of exotic spacetimes with CTCs.
\\
Summing up, the spacetime $\TT$ defined in Eq. \eqref{TTdef} possesses no curvature singularity and contains CTCs, besides being manifestly topologically trivial. The forthcoming sections describe other noteworthy features of $\TT$, mostly retracing the analysis detailed in \cite{CQG}.
\vspace{0.2cm}\\
\textbf{A case study.}
Following \cite{CQG}, as a prototype for numerical computations in this work we systematically refer to the setting specified hereafter:
\begin{equation}\begin{array}{c}
\dd{\ellm = {3 \over 5}\,R\,, \quad \ellp = {4 \over 5}\,R\,, \quad a = {9 \over 100}\,, \quad b = 10\,, \quad 
\HH(y) := \mathfrak{H}\!\left({\ellp/\!R - y \over \ellp/\!R - \ellm/\!R}\right) ;}\vspace{0.2cm}\\
\dd{\mathfrak{H}(w) := \!\left\{\!\!\begin{array}{ll}
\dd{0} & \dd{\mbox{for $w < 0$}\,,} \vspace{0.cm} \\
\dd{35\, w^4 (1\!-\!w)^3\! + 21\, w^5 (1\!-\!w)^2\! + 7\, w^6 (1\!-\!w)\! + \!w^7} &
\dd{\mbox{for $0 \leqs w \leqs 1$}\,,} \vspace{0.cm} \\
\dd{1} & \dd{\mbox{for $w > 1$}\,.} 
\end{array}\right.} \label{cho1}
\end{array}\end{equation}
For $\HH$ as above, the shape function $\XX$ is of class $C^3$. The choices indicated in Eq. \eqref{cho1} are not completely arbitrary, but partly motivated by some requirements to be discussed in Section \ref{ffSec} (see Eq. \eqref{impl}). The parameter $R$, intentionally left unspecified, plays the role of natural length scale for the model under analysis.
\newpage

\section{Natural choices of time and space orientations}\label{SecOr}
\vspace{-0.15cm}
Consider the following vector fields:
\begin{equation}\begin{array}{c}
\dd{E_{(0)} :=\, {(1 - \XX)\,\rho\,\de_t
+ \XX\, b\,\de_\ff \over \rho\,(1 - \XX)^2\!
+ a\,b\,R\,\XX^2} ~, \qquad
E_{(1)} :=\, {(1 - \XX)\,\de_\ff - \XX a R\,\de_t
\over \rho\,(1 - \XX)^2\! + a\,b\,R \,\XX^2} ~,} \vspace{0.2cm} \\
\dd{E_{(2)} :=\, \de_\rho ~, \qquad E_{(3)} :=\, \de_z ~.} \label{EComp}
\end{array}\end{equation}
These are everywhere defined on $\R^4$ and have the same regularity of $\XX$ (i.e., are of class $C^k$); furthermore, it can be checked by explicit computations (see \cite{CQG}) that
\begin{equation}
g(E_{(\al)},E_{(\be)}) = \eta_{\al \be} \qquad \mbox{{\small $(\al,\be \!\in\! \{0,1,2,3\})$}}\,, \label{normE}
\end{equation}
where $(\eta_{\al \be}) \!:=\! \mbox{diag}(-1,1,1,1)$. In other words, the set $(E_{(\alpha)})_{\alpha \in \{0,1,2,3\}}$ forms a pseudo-orthonormal tetrad; this can be used to induce natural choices of time and space orientations for $\TT$, following the prescriptions outlined in the sequel.
\\
Notably, $E_{(0)}$ is everywhere timelike (see Eq. \eqref{normE}). Taking this into account, we fix the time orientation of $\TT$ assuming that $E_{(0)}$ is future-directed at all points. This time orientation reproduces the familiar Minkowskian one in the spacetime region outside $\To_{\ellp}$, where $E_{(0)}\! = \de_t$\,. At the same time, the established convention implies that $\de_\ff$ is timelike and future-directed inside $\To_{\ellm}$, since $E_{(0)}\! = 1/(a R)\,\de_\ff$ therein. Of course, the integral curves of $E_{(0)}$ can be interpreted as the worldlines of certain \emph{fundamental observers}; any event $p \in \TT$ uniquely identifies one of these observers and $E_{(0)}$ coincides with the $4$-velocity of the latter.
\\
Next, consider the vector fields $(E_{(i)})_{i \in \{1,2,3\}}$; these are everywhere spacelike (see Eq. \eqref{normE}) and further fulfil $E_{(1)}\! = (1/\rho)\,\de_\ff$, $E_{(2)}\! = \de_\rho$, $E_{(3)}\! = \de_z$ in the Minkowskian region outside $\To_{\ellp}$. In view of these facts, we choose to equip $\TT$ with the space orientation induced by the left-handed ordered triplet $(E_{(1)},E_{(2)},E_{(3)})$.
\\
Finally, let us mention two additional facts regarding the tetrad $(E_{(\alpha)})_{\alpha \in \{0,1,2,3\}}$. On the one hand, the family $(E_{(i)})_{i \in \{1,2,3\}}$ is not involutive; by Frobenius theorem, this means that there does not exists a foliation of $\TT$ into spacelike hypersurfaces orthogonal to $E_{(0)}$. On the other hand, none of the vector fields $(E_{(\alpha)})_{\alpha \in \{0,1,2,3\}}$ fulfils the Killing equation $\mathcal{L}_{E_{(\alpha)}}g = 0$ ($\mathcal{L}$ indicates the Lie derivative); so, none of them generates an isometry of $\TT$.

\section{Manifest symmetries and stationary limit surfaces}\label{SecSym}
\vspace{-0.15cm}
The spacetime under analysis exhibits a number of symmetries; hereafter these symmetries are discussed, together with some of their most relevant implications (see \cite[Sec. 4]{CQG} for more details on this theme).
\vspace{0.1cm}\\
First of all, notice that the line element \eqref{dsXidef} is invariant under each of the following coordinate transformations, describing \emph{discrete symmetries} of the spacetime $\TT$:
\begin{eqnarray}
(t,\ff,\rho,z) \to (-t,-\ff,\rho,z)\,; \label{Sym1} \\
(t,\ff,\rho,z) \to (t,\ff,\rho,-z)\,. \label{Sym2}
\end{eqnarray}
Incidentally, it should be observed that the symmetry by reflection across the plane $\{z = 0\}$ represented by Eq. \eqref{Sym2} depends crucially on the distinguished choice \eqref{Xidef} of the shape function $\XX$ (granting that $\XX(\rho,z) = \XX(\rho,-z)$). 
\\
In terms of the orthonormal tetrad introduced in Section \ref{SecOr} (see Eq. \eqref{EComp}), the above transformations \eqref{Sym1}\! \eqref{Sym2} induce, respectively, the mappings $(E_{(0)}, E_{(1)},E_{(2)},E_{(3)}) \to (- E_{(0)},- E_{(1)}, E_{(2)}, E_{(3)})$ and $(E_{(0)},E_{(1)},E_{(2)},E_{(3)}) \to (E_{(0)}, E_{(1)}, E_{(2)}, -E_{(3)})$. The facts mentioned above and the considerations of Section \ref{SecOr} indicate, in particular, that the spacetime $\TT$ is invariant under the (separate) inversion of time and space orientations.
\vspace{0.1cm}\\
To proceed, let us examine the presence of (global) \emph{continuous symmetries}. In this connection, it should be noticed that the coefficients of the metric $g$ do not depend explicitly on $t,\ff$ (recall that $\XX \equiv \XX(\rho,z)$); this suffices to infer that the following are Killing vector fields, generating each a one-parameter group of isometries of $g$:
\begin{equation}
K_{(0)} := \de_t\,, \qquad K_{(1)} := \de_\ff\,. \label{K0K1}
\end{equation}
Next, consider the related set of events
\begin{equation}
\Sigma_{(i)} := \{p \in \TT\;|\;g_p(K_{(i)},K_{(i)}) = 0 \} \qquad (i = 0,1)\,,
\end{equation}
admitting the explicit coordinate representations
\begin{equation}
\Sigma_{(0)} = \{\XX(\rho,z) = (1+b)^{-1} \}\,, \qquad 
\Sigma_{(1)} = \{\XX(\rho,z) = (1+a R/\rho)^{-1} \}\,. \label{Sigma}
\end{equation}
Both $\Sigma_{(0)}$ and $\Sigma_{(1)}$ are timelike hypersurfaces contained in the spacetime region delimited by the tori $\To_{\ellm},\To_{\ellp}$ (see Fig. 3); in particular, neither of them is a Killing horizon.
As a matter of fact, $\Sigma_{(0)}$ and $\Sigma_{(1)}$ are \emph{stationary limit surfaces} \cite{Hawk} for $K_{(0)}$ and $K_{(1)}$, respectively. This means that the integral curves of $K_{(0)}$ can be interpreted as the worldlines of stationary observers outside $\Sigma_{(0)}$ (where $\XX(\rho,z) \!>\! (1+b)^{-1}$ and $K_{(0)}$ is timelike), but no particle travelling along a timelike curve inside $\Sigma_{(0)}$ (where $\XX(\rho,z) \!<\! (1+b)^{-1}$ and $K_{(0)}$ is spacelike) can remain at rest with respect to such observers. Similarly, no particle outside $\Sigma_{(1)}$ can remain at rest with respect to observers moving along the orbits of $K_{(1)}$ inside $\Sigma_{(1)}$.
\begin{figure}[t!]\label{fig:HorPot}
\vspace{-0.7cm}
    \centering
        \begin{subfigure}[b]{0.475\textwidth}\label{fig:Hor}
        		\centering
                \includegraphics[width=5cm]{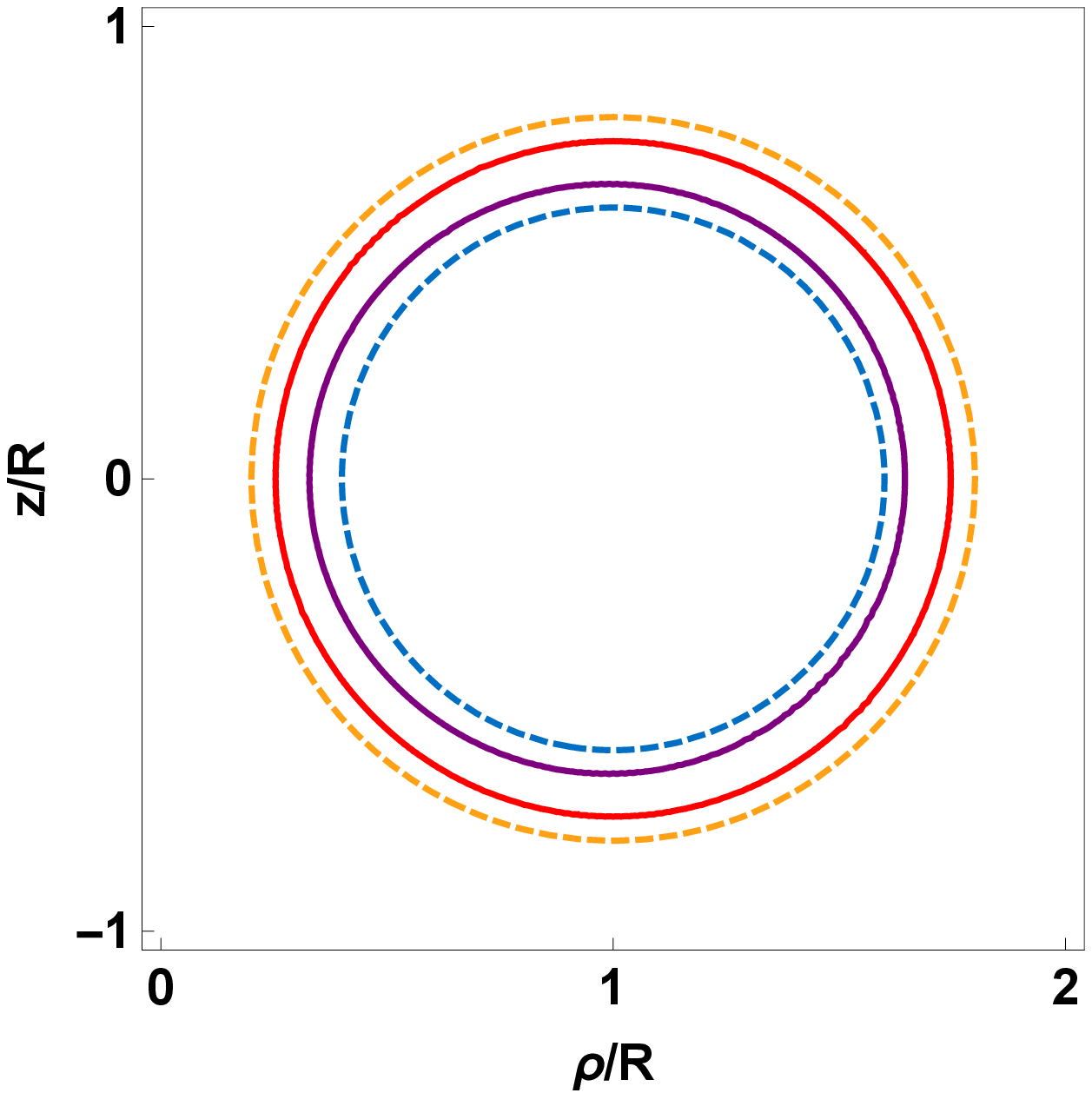}
                \caption*{\textsc{Figure 3.} Representation (for fixed $t,\ff$) of the stationary limit surfaces 
                $\Sigma_{(0)}$ (in red), $\Sigma_{(1)}$ (in purple), along with the tori $\To_{\ellm}$ (in blue),
                $\To_{\ellp}$ (in orange) (cf. Eq.s \eqref{cho1} \eqref{Sigma}).}
        \end{subfigure}
        \hspace{0.3cm}
        \begin{subfigure}[b]{0.475\textwidth}
                \includegraphics[width=\textwidth]{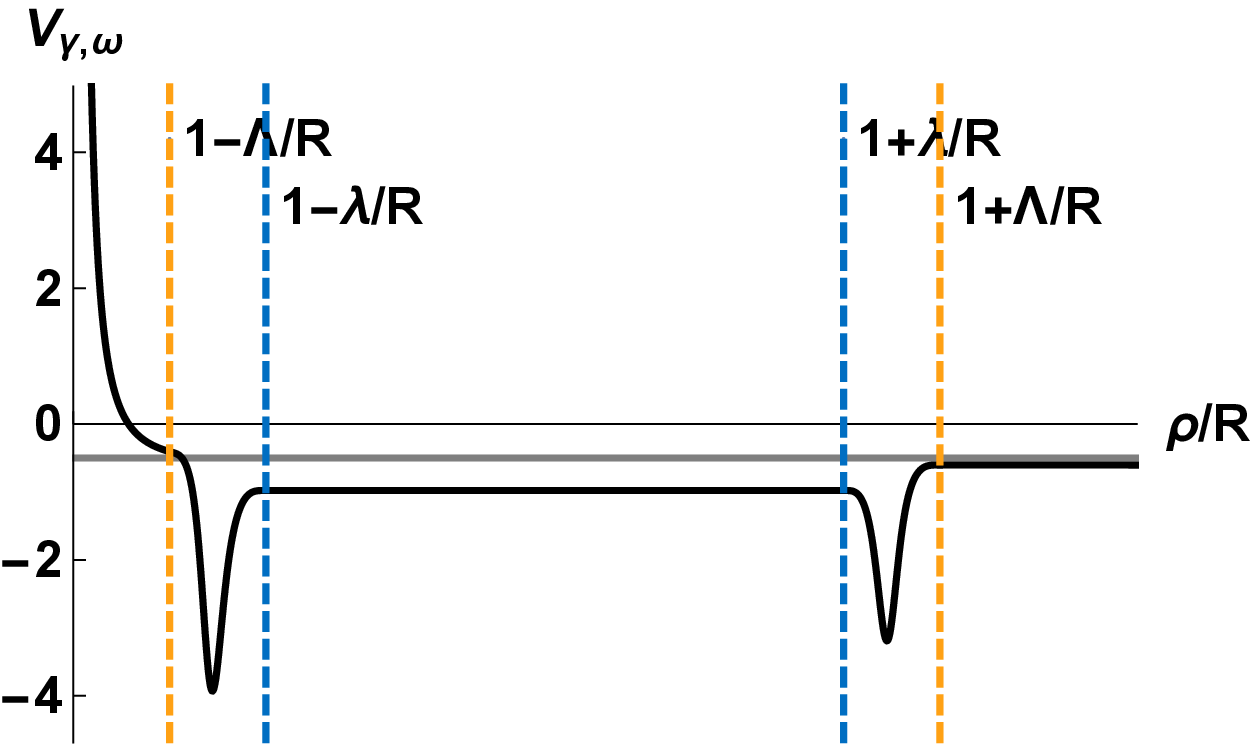}\vspace{0.4cm}\label{fig:Pot}
                \caption*{\textsc{Figure 4.} Plot of the potential $V_{\ga,\om}$ for $\ga = 1.1$, $\om = 0.08$
                (cf. Eq.s \eqref{cho1} \eqref{Vdef}). The gray line represents the energy level $E= -1/2$ (cf. Eq.s
                \eqref{LEdef} \eqref{rhoQuad}).\\
				\vspace{-0.5cm}
                }
        \end{subfigure}
\end{figure}

\section{Quadratures for a special class of geodesics}\label{quadSec}
\vspace{-0.15cm}
Consider the Lagrangian function
\begin{equation}
L(U) := {1 \over 2}\,g(U,U) \qquad (U \!\in\! T\TT) \label{Lag}
\end{equation}
and recall that solutions of the associated Euler-Lagrange equations describe affinely parametrized geodesics. It should also be kept in mind that for any geodesic $\xi$ with tangent vector field $\dot{\xi}$, modulo possible re-parametrizations, there holds
\begin{equation}
L(\dot{\xi}) = \mbox{const.} = E\,, \qquad E = \left\{\!\!\begin{array}{ll}
\dd{- 1/2} & \dd{\mbox{if $\xi$ is timelike}\,,} \\
\dd{0} & \dd{\mbox{if $\xi$ is null}\,,} \\
\dd{+ 1/2} & \dd{\mbox{if $\xi$ is spacelike}\,.} \\
\end{array}\right. \label{LEdef}
\end{equation}
This fact and the symmetries of $\TT$ discussed in Section \ref{SecSym} allow to reduce the Euler-Lagrange equations to quadratures, at least for a special class of geodesics. 
More precisely, in the following we restrict the attention to geodesics lying in the plane $\{z = 0\}$, and essentially retrace the analysis of \cite[Sec. 5]{CQG}. \\
On account of the reflection symmetry \eqref{Sym2}, the geodesics mentioned above can be analysed considering just the restriction of the Lagrangian \eqref{Lag} to the cotangent bundle of the plane $\{z\!=\! 0\}$; this has the following coordinate representation, where $\XX \equiv \XX(\rho,0) \equiv \HH(|\rho/R - 1|)$ (see Eq. \eqref{Xidef}):
\begin{equation}
L(\dot{\xi})\big|_{z = 0,\dot{z} = 0} \!= -\,{1 \over 2} \big[(1\!-\!\XX)\,\dot{t} + a R\, \XX \dot{\ff}\big]^2\!
+ {1 \over 2} \big[\rho\, (1\!-\!\XX) \dot{\ff} - b\,\XX \dot{t}\,\big]^2\!
+ {1 \over 2}\, \dot{\rho}^2 \,. \label{redL}
\end{equation}
The above reduced Lagrangian possesses a maximal number of first integrals: namely, the canonical momenta associated to the cyclic coordinates $t,\ff$ (coinciding with the Noether charges related to the Killing vector fields $K_{(0)},K_{(1)}$ of Eq. \eqref{K0K1}) and the energy of the Lagrangian system.\\
In place of the above mentioned canonical momenta, it is convenient to consider the related dimensionless parameters
\vspace{-0.15cm}
\begin{equation}
\ga := -\,{\de L(\dot{\xi})|_{z = 0,\dot{z} = 0} \over \de \dot{t}}\,, \qquad 
\om := -\,{1 \over \ga\,R}\,{\de L(\dot{\xi})|_{z = 0,\dot{z} = 0} \over \de \dot{\ff}}\,. \label{gaom}
\end{equation}
It can be checked by explicit computations that the above definitions yield
\vspace{-0.1cm}
\begin{eqnarray}
\dot{t} = \ga\,, \qquad \dot{\ff} = -\,{\ga\,\om\,R \over \rho^2} \qquad\quad
\mbox{outside $\To_{\ellp}$}\,, \label{tdfd0} \vspace{-0.15cm}\\
\dot{t} = -\,{\ga \over b^2}\,, \qquad \dot{\ff} = {\ga\,\om \over a^2 R} \qquad\quad
\mbox{inside $\To_{\ellm}$}\,. \label{tdfd1}
\end{eqnarray}
Eq.s \eqref{tdfd0} \eqref{tdfd1} allow to infer a number of notable facts.
\vspace{-0.15cm}
\begin{enumerate}[(i)]
\item Causal geodesics traversing the Minkowskian region outside $\To_{\ellp}$ are future-directed (recall the time orientation established in Section \ref{SecOr}) only for $\ga > 0$. In particular, for timelike curves of this kind, $\ga \equiv dt/d\tau$ is the usual Lorentz factor of special relativity; so, $\ga \!\geqs\! 1$ and the limits $\ga \!\to\!1^+\!$, $\ga \!\to\! +\infty$ correspond, respectively, to non-relativistic and ultra-relativistic regimes.
\vspace{-0.15cm}
\item Causal geodesics travelling across the flat region inside $\To_{\ellm}$ are future-directed only if $\ga \cdot \om > 0$.\vspace{-0.15cm}
\item In view of the previous remarks (i)(ii), causal geodesics crossing both the region outside $\To_{\ellp}$ and the one inside $\To_{\ellm}$ are future-directed only if
\vspace{-0.05cm}
\begin{equation}
\ga > 0 \qquad \mbox{and} \qquad \om > 0\,. \vspace{-0.05cm}
\end{equation}
Remarkably, for $\ga > 0$ the first identity in Eq. \eqref{tdfd1} gives $\dot{t} < 0$ inside $\To_{\ellm}$, indicating that the coordinate $t$ decreases while moving towards the future. One can make sense of the latter (apparently paradoxical) feature keeping in mind that $t$ is a spatial coordinate inside $\To_{\ellm}$; this fact plays a crucial role in the construction of time travels, to be discussed in the upcoming Section \ref{ffSec}.
\end{enumerate}
To proceed, notice that the conservation of the energy associated to the reduced Lagrangian \eqref{redL} can be expressed, in terms the parameters $\ga,\om$ via the relation
\begin{equation}
{1 \over 2}\,\dot{\rho}^2 + V_{\ga,\om}(\rho) = E\,, \label{rhoQuad}
\end{equation}
where $E$ is a constant fixed as in Eq. \eqref{LEdef} and ($\XX \equiv \XX(\rho,0) \equiv \HH(|\rho/\!R - 1|)$)
\begin{equation}
V_{\ga,\om}(\rho) := \ga^2 \, {[a\,\XX - (1\!-\!\XX)\,\om]^2\!
- [(\rho/\!R)\,(1\!-\!\XX) + b\,\XX\,\om]^2 \over
2\,[(\rho/\!R)\,(1\!-\!\XX)^2 + a\,b\,\XX^2]^2} \;. \label{Vdef}
\end{equation}
\newpage
\noindent
In particular, there holds
\begin{eqnarray}
V_{\ga,\om}(\rho) = {\ga^2 \over 2} \left({R^2 \om^2 \over \rho^2} - 1 \right) 
\qquad \mbox{outside $\To_{\ellp}$}\,, \label{VXRed0} \vspace{-0.1cm} \\
V_{\ga,\om}(\rho) = \,\mbox{const.} = {\ga^2 \over 2 a^2}\left({a^2 \over b^2} - \om^2\right)
\qquad \mbox{inside $\To_{\ellm}$} \,. \label{VXRed1}
\end{eqnarray}
At the same time, $V_{\ga,\om}$ depends sensibly on the sign of $\om$ in the intermediate region delimited by $\To_{\ellm}$ and $\To_{\ellp}$. Fig. 4 shows a plot of $V_{\ga,\om}$ in the setting \eqref{cho1}, for a particular choice of $\ga,\om$ (specified in the caption).\\
Eq. \eqref{rhoQuad} and the definitions of Eq. \eqref{gaom} allow to derive, by standard methods, explicit quadrature formulas for the solutions $t(\tau),\ff(\tau),\rho(\tau)$ ($\tau \!\in\! \R$) of the Euler-Lagrange equations associated to the reduced Lagrangian \eqref{redL}; for brevity, we do not report these formulas here and refer to \cite{CQG} for more details.

\section{Exotic causal geodesics and time travel by free fall}\label{ffSec}
\vspace{-0.15cm}
Building on the results described in the previous section, it can be shown that there exist future-directed causal geodesics starting at a point in the Minkowskian region outside $\To_{\ellp}$, reaching the flat region inside $\To_{\ellm}$ and finally returning outside $\To_{\ellp}$, arriving at an event in the causal past of the initial point. Such geodesics can be unequivocally interpreted as the trajectories of (massive or massless) test particles performing a free fall time travel into the past, with respect to observers at rest in the outer Minkowskian region. Using a suggestive terminology, the latter observers could refer to the spacetime region delimited by $\To_{\ellp}$ as a ``time machine''.\\
As anticipated in item (iii) of the preceding section, it should be kept in mind that the existence of geodesics such as those described above relies significantly on the possibility to implement the condition $\dot{t} < 0$ inside $\To_{\ellm}$. Understandably, this region can be reached only if the parameters $a,b,\ellm,\ellp,R$ and the shape function $\XX$ characterizing the spacetime $\TT$, as well as the initial data related to the geodesic coefficients $\ga,\om$, are chosen suitably (in this connection, see Eq.\! \eqref{ass} below).\\
Let $\xi$ be a causal geodesic with initial data $\xi(0), \dot{\xi}(0)$, described in coordinates by
\begin{equation}\begin{array}{llll}
\dd{t\ze = 0\,,}\;	& \dd{\ff\ze = 0\,,}\; 	& \dd{\rho\ze = \rho_0 \!>\! R \!+\! \ellp\,,}\;	& \dd{z\ze = 0\,;} \vspace{0.1cm} \\
\dd{\dot{t}\ze = \ga \!>\! 0\,,}\; 	& \dd{\dot{\ff}\ze = - \ga\,\om R/\rho_0^2 \!<\! 0\,,}\;	& \dd{\dot{\rho}\ze < 0\,,}\;	& \dd{\dot{z}\ze = 0\,.} \label{data}
\end{array}\end{equation}
In the above positions, $t\ze$ and $\ff\ze$ are fixed arbitrarily, while the choices of $\rho\ze$ and $z\ze$ indicate that the initial event lies outside $\To_{\ellp}$, on the plane \!$\{z\! =\! 0\}$. On the other hand, $\dot{t}\ze \!=\! \ga \!>\! 0$ grants that $\xi$ is future-directed (see item (i) of Section \ref{quadSec}), whereas $\dot{\rho}\ze \!<\! 0$ implies that $\xi$ is initially moving towards the torus $\To_{\ellp}$. The specific expression for $\dot{\ff}\ze$ follows essentially from the definition of $\om$ given in Eq. \eqref{gaom}; note that it understands the assumption $\om \!>\! 0$, ensuring that $\xi$ remains future-directed when it enters the region delimited by $\To_{\ellm}$. Finally, notice that the results of Section \ref{quadSec} can certainly be employed, since $z\ze = 0$,\! $\dot{z}\ze = 0$.
\\
A sufficient condition granting that $\xi$ crosses the region inside $\To_{\ellm}$ is the following:
\begin{equation}
\mbox{$\exists\,\rho_{1} \in (0,R - \ellm)$\; s.t.\; $V_{\ga,\om}(\rho) \leqs E$\, for all\, $\rho \geqs \rho_1$}\,, \label{ass}
\end{equation}
where $V_{\ga,\om}$ and $E$ are defined, respectively, as in Eq.s \eqref{Vdef} and \eqref{LEdef}.
Especially, it should be kept in mind that $E = -1/2$ if $\xi$ is timelike, and $E = 0$ if $\xi$ is null.\\
In view of Eq.s \eqref{VXRed0} \eqref{VXRed1}, to fulfil the condition \eqref{ass} it is necessary to comply at least with the requirements
\begin{equation}\begin{array}{c}
\dd{{a \over b} < 1 - {\ellp \over R}\,, \qquad 
\ga > \sqrt{|2E|}\,\sqrt{{(1\!-\! \ellp/R)^2 + a^2 \over (1 \!-\! \ellp/R)^2 - a^2/b^2}}\;,} \vspace{0.1cm}\\
\dd{{a \over b}\,\sqrt{1 - {2 E \,b^2 \over \ga^2}} < \om <
\left(1 - {\ellp \over R} \right) \sqrt{1 + {2E \over \ga^2}}\,.}  \label{impl}
\end{array}\end{equation}
\newpage
\noindent
This shows that a carefully devised choice of parameters and initial data is required. For example, radial motions starting outside $\To_{\ellp}$ with $\dot{\ff} = 0$ cannot enter the spacetime region delimited by $\To_{\ellm}$, since the corresponding momentum $\om = 0$ (see Eq. \eqref{tdfd0}) does not fulfil the above Eq. \eqref{impl}.\\
Incidentally, the specific choices reported in Eq. \eqref{cho1} do satisfy the condition \eqref{ass} for any $\ga,\om$ fulfilling the requirements in Eq. \eqref{impl}, for both $E = -1/2$ (timelike geodesics) and $E = 0$ (null geodesics).
\\
Whenever the condition \eqref{ass} is fulfilled, the radial variable $\rho(\tau)$ associated to the causal geodesic $\xi$ starts at $\rho(0) = \rho_0 > R + \ellp$ (see Eq. \eqref{data}), decreases until it reaches the value $\rho_1$ (defined implicitly by Eq. \eqref{ass}) and then begins to increase infinitely. Especially, $\rho(\tau)$ returns to its initial value at some proper time $\tau_2$, i.e.,
\begin{equation}
\rho(\tau_2) = \rho_0\,; \label{ro2}
\end{equation}
in addition, it is possible to arrange things so that
\begin{equation}
t(\tau_2) < 0\,, \qquad \ff(\tau_2) = 0\, (\mbox{mod $2\pi$})\,. \label{t2f2}
\end{equation}
The above relations indicate that the geodesic $\xi$ returns at exactly the same spatial position from which it started off, but at an earlier coordinate time. Thus, a test particle freely falling along $\xi$ would be able to reach an event lying in the causal past of its departure, even in the outer Minkowskian region.\\
Using the quadrature formulas mentioned in Section \ref{quadSec}, it is possible to characterize quantitatively the geodesic motion described above, both for timelike and null curves. As an example, consider a timelike geodesic $\xi$ with 
\begin{equation}
E = -1/2\;,
\end{equation}
fulfilling the conditions \eqref{data} \eqref{ro2} \eqref{t2f2}. An exhaustive analysis of this case is provided in \cite[Sec. 6]{CQG}; in particular, it is shown that in the limit (cf. Eq. \eqref{impl})
\begin{equation}
\xf \,:=\, \sqrt{\ga^2 - {a^2 \over b^2} \left(1 + {b^2 \over \ga^2}\right)} \;\to\; 0^+ \label{defxf}
\end{equation}
there holds (at least for suitable choices of the shape function $\HH$; see \cite{CQG})
\begin{equation}\begin{array}{c}
\dd{{t(\tau_2) \over R} = - \left({4 a \over b^2}\,{\ellm \over R} \right)\!{1 \over \xf}\,
\big(1 \!+\! o(1)\big) \,, \qquad\!
{\tau_2 \over R} = \!\left({4 a \over \ga}\,{\ellm \over R}\right) {1 \over \xf}\;
\big(1 \!+\! o(1)\big)\,,} \vspace{0.2cm}\\
\dd{\ff(\tau_2) = \left({4 \over b}\,{\ellm \over R}\,\sqrt{1 + {b^2 \over \ga^2}}\,\right)\! {1 \over \xf}\,
\big(1 \!+\! o(1)\big)\;\;(\mbox{mod $2\pi$}) ~.}
\end{array}\end{equation}
A few remarks regarding the above asymptotic expansions are in order:
\vspace{-0.15cm}
\begin{enumerate}[(i)]
\item $t(\tau_2)$ can be large and negative at will, showing that the amount of coordinate time travelled into the past can be made as large as desired.
\vspace{-0.15cm}
\item $\tau_2$ becomes awfully large as well for $\xf \!\to\! 0^+$. However, since $\tau_2/|t(\tau_2)| \!\sim b^2\!/\ga$, one has that $\tau_2$ remains comparatively small with respect to $|t(\tau_2)|$ at least for ultra-relativistic motions (i.e., for large $\ga$); this shows that the amount of proper time spent during the time travel can be kept small by moving at sufficiently high speeds, exactly as in special relativity.
\vspace{-0.15cm}
\item $\ff(\tau_2)$ varies quickly for $\xf \!\to\! 0^+$, indicating that the identity $\ff(\tau_2) \!=\! 0$ $(\mbox{mod $2\pi$})$ is always allowed in principle, but difficult to achieve in practice since it relies on a fine tuning of the parameters. Of course, this fine tuning cannot be described using just the above asymptotic expansion; on the other hand, let us suggest that it could be avoided by considering a piecewise geodesic motion (see Section \ref{secfut} for further comments of this topic).
\end{enumerate}
To conclude this section let us report a few numerical results extracted from Tables 1-3 of \cite{CQG}, where many more examples can be found. For definiteness, we refer to the specific setting \eqref{cho1} and consider a massive particle starting in the outer Minkowskian region with a speed comparable to those attained at the Large Electron-Positron Collider; namely, we set
\begin{equation}
\ga = 10^5\,. \label{gaNum}
\end{equation}
As an example, we require the said particle to perform a time travel by free fall so as to reach an event lying roughly a thousand years in the past, while experiencing an interval of proper time of only about one year:
\begin{equation}
t(\tau_2) \sim -\, 10^3\,\ye\,, \qquad \tau_2 \sim 1 \,\ye\,.
\end{equation}
To this purpose it is necessary to fix suitably the parameter $\xf$ of Eq. \eqref{defxf}, depending on the size $R$ of the apparatus. The forthcoming Eq.s \eqref{eq1}-\eqref{eq3} indicate the required choices of $\xf$ for three prototypical values of $R$, corresponding, respectively, to a nominal human-sized length ($10^2 \me$), the Sun-Earth distance ($10^{11} \me$) and to an inter-stellar distance of a hundred light-years ($10^{18} \me$):
\begin{eqnarray}
\xf \sim 2 \!\cdot\! 10^{-20} \qquad \mbox{for $R = 10^2 \me$}\,; \label{eq1}\\
\xf \sim 2\!\cdot\!10^{-11} \qquad \mbox{for $R = 10^{11} \me$}\,; \label{eq2}\\
\xf \sim 2\!\cdot\!10^{-4} \qquad \mbox{for $R = 10^{18} \me$}\,. \label{eq3}
\end{eqnarray}
In all the above cases it is possible to fine tune $\xf$ so as to implement the condition $\ff(\tau_2) = 0\,\mbox{(mod $2\pi$)}$, but we omit the discussion of this issue for brevity (see \cite{CQG}).

\section{Tidal forces during time travel by free fall}\label{secTid}
\vspace{-0.15cm}
Consider a small extended body performing a time travel by free fall into the past. More precisely, assume that the body consists of nearby massive point particles whose worldlines form a beam of timelike geodesics of the type described in Section \ref{ffSec}; this understands, in particular, that the interaction between different particles is being neglected. Hereafter the tidal forces experienced by the said extended body (i.e., the relative accelerations of its constituent particles) are investigated.\\
Let $\xi$ be the geodesic followed by one of the particles in the body and let $\delta\xi$ be the deviation vector from it \cite{Wald}.
As well-known, tidal accelerations between $\xi$ and nearby geodesics are described by the Jacobi deviation equation
\begin{equation}
{\nabla^2 \delta\xi \over d \tau^2}(\tau) = \AAt\, \delta \xi(\tau)\,, \qquad 
\AAt X \!:= -\,\Riem\big(X, \dot{\xi}(\tau)\big)\, \dot{\xi}(\tau)\,,
\end{equation}
where $\Riem$ is the Riemann tensor corresponding to the metric $g$.
For any fixed $\tau$, the orthogonal complement of $\dot{\xi}(\tau)$ is a Euclidean subspace of $T_{\xi(\tau)}\TT$ and $\AAt$ determines a self-adjoint operator on it. The \emph{maximal tidal acceleration per unit length} at $\xi(\tau)$ is defined as (see \cite[Sec. 7]{CQG} for further details)
\begin{equation}
\aa(\tau) \,:= \sup_{\mbox{{\small $X \!\in\! T_{\xi(\tau)}\TT \backslash \{0\}$, $g(\dot{\xi}(\tau),X)\! =\! 0$}}}
\!{\sqrt{g(\AAt X, \AAt X)} \over \sqrt{g(X,X)}}~. \label{defaa}
\end{equation}
It can be shown by direct inspection that
\begin{equation}
\aa(\tau) = {\ga^2 \over R^2}\;\ac \big(\rho(\tau)/\!R\big) \,, \label{atau}
\end{equation}
where $\ac(\rho/\!R)$ is a dimensionless function, depending on the spacetime parameters $\ellm/R,\ellp/R,a,b$ and on the geodesic coefficients $\ga,\xf$. Of course, $\ac(\rho/\!R) = 0$ for $\rho \in (0,R - \ellp) \cup (R - \ellm,R+ \ellm) \cup (R+\ellp,+\infty)$, correctly indicating that tidal accelerations vanish in the flat spacetime regions outside $\To_{\ellp}$ and inside $\To_{\ellm}$.
\newpage
\noindent
\begin{figure}[t!]\label{fig:acZif}
\vspace{-0.7cm}
    \centering
        \begin{subfigure}[b]{0.475\textwidth}\label{fig:ac}
                \includegraphics[width=\textwidth]{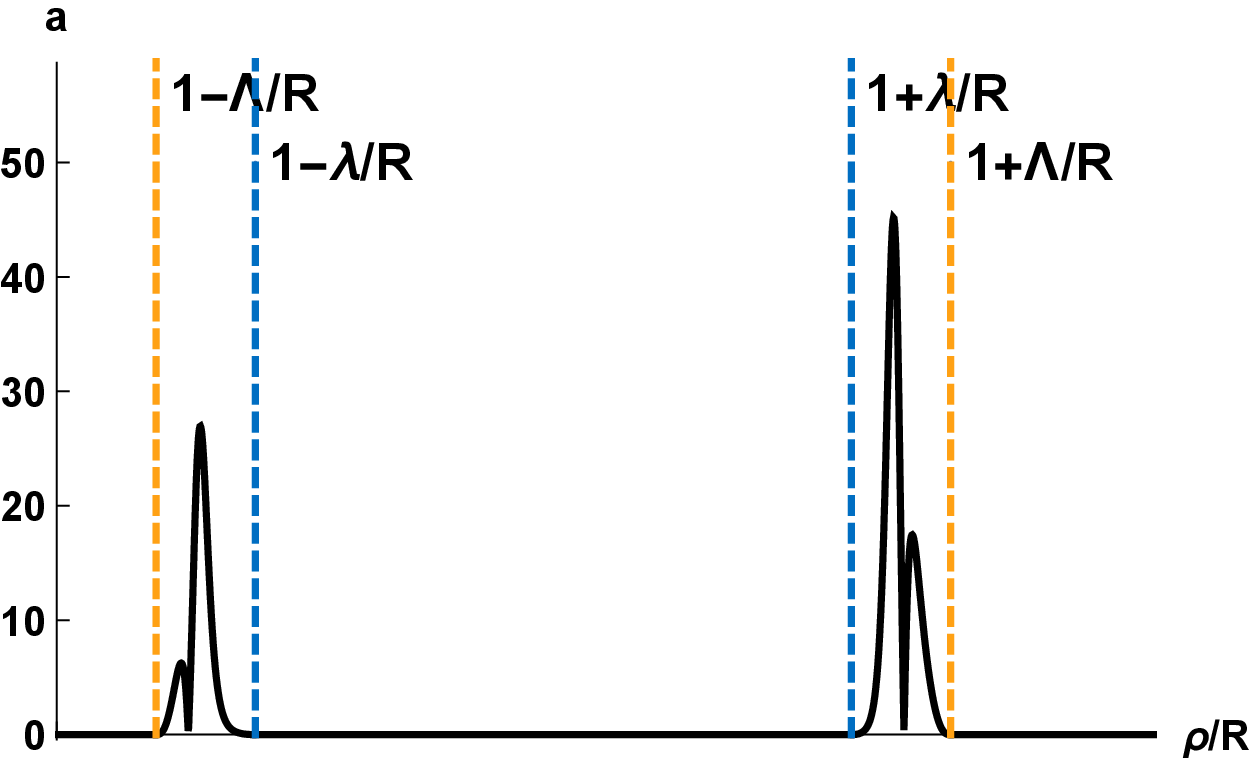}\vspace{-0.cm}
                \caption*{\textsc{Figure 5.} Plot of the tidal acceleration function $\ac(\rho/\!R)$ (cf. Eq.s \eqref{cho1} 						\eqref{atau}) for $\ga = 10^5$, $\xf = 2 \cdot 10^{-20}$.}
        \end{subfigure}
        \hspace{0.4cm}
        \begin{subfigure}[b]{0.475\textwidth}
                \includegraphics[width=\textwidth]{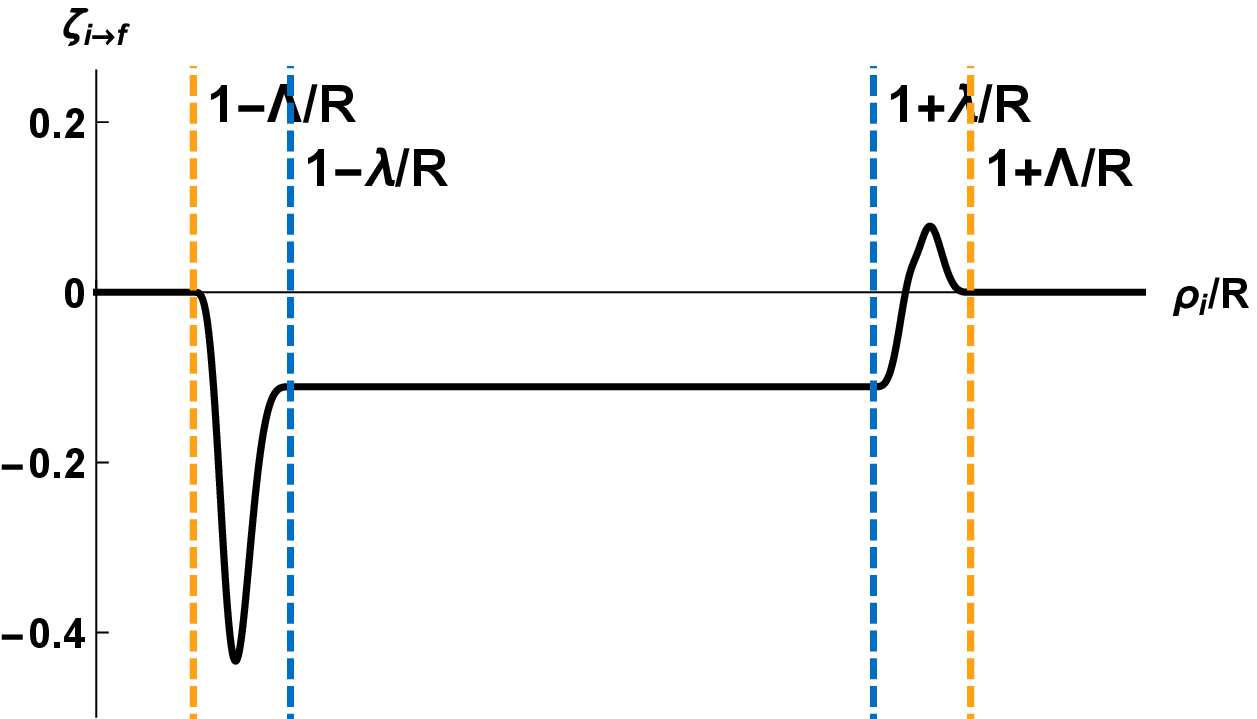}\label{fig:Zif}
                \caption*{\textsc{Figure 6.} Plot of the redshift factor $\zeta_{i \to f}$ (cf. Eq.s \eqref{cho1} \eqref{zif})						for $\om = 0.08$, as a function of the emission coordinate $\rho_i/\!R$.}
        \end{subfigure}
\end{figure}
As an example, refer again to the setting \eqref{cho1} and consider a geodesic motion with $\ga = 10^5$ (see Eq. \eqref{gaNum}). Fig. 5 shows the plot of $\ac(\rho/\!R)$ in this case, for $\xf = 2 \cdot 10^{-20}$ (see Eq. \eqref{eq1}).
The following numerical results are extracted form Tables 1-3 of \cite{CQG} ($\gt\! = 9.8\, \me/\se^2$ is a nominal value for the Earth's gravitational acceleration):
\begin{eqnarray}
\max_{\tau \in [0,\tau_2]} \aa(\tau) \sim 4 \cdot 10^{23}\, \gt/\me \qquad \mbox{for $R = 10^2 \me$\,, $\xf = 10^{-20}$}\,; \label{aq1}\vspace{-0.1cm}\\
\max_{\tau \in [0,\tau_2]} \aa(\tau) \sim 4 \cdot 10^{5}\, \gt/\me \qquad \mbox{for $R = 10^{11} \me$\,, $\xf = 10^{-11}$}\,; \label{aq2}\vspace{-0.1cm}\\
\max_{\tau \in [0,\tau_2]} \aa(\tau) \sim 4 \cdot 10^{-9}\, \gt/\me \qquad \mbox{for $R = 10^{18} \me$\,, $\xf = 10^{-4}$}\,. \label{aq3}
\end{eqnarray}
The above numerics show that, for the chosen value of $\ga$, tidal accelerations are intolerable for a human being if the radius $R$ is smaller than the Sun-Earth distance, while they become essentially negligible for a radius $R$ of galactic size.

\section{Redshift and blueshift of light signals}\label{RedSec}
\vspace{-0.15cm}
The existence of future-directed causal geodesics lying in the plane $\{z = 0\}$ and connecting the spacetime region outside $\To_{\ellp}$ to the one inside $\To_{\ellm}$ was discussed in Section \ref{ffSec}. In the following, working in the geometric optics approximation, it is analysed the frequency of light signals propagating along null geodesics of the above type, as measured by suitable global observers.\\
More precisely, consider the fundamental observers described in Section \ref{SecOr} and recall that their worldlines are the integral curves of the timelike vector field $E_{(0)}$, belonging to the tetrad \eqref{EComp}. The frequency of a light signal propagating along a null geodesic $\xi$ (with tangent $\dot{\xi}$), measured at $p \in \TT$ by a fundamental observer is
\begin{equation}
\Omega := -\,g_p(\dot{\xi},E_{(0)})\,. \label{freq}
\end{equation}
Let a light signal of frequency $\Omega_i$ be emitted at a spacetime point $p_i \in \TT$ by an observer of the above family and assume it to reach an event $p_f$ where another fundamental observer measures the frequency $\Omega_f$; the conventional \emph{redshift factor} associated to this scenario is
\begin{equation}
\zeta_{i \to f} := {\Omega_i \over \Omega_f} - 1\,. \label{redshift}
\end{equation}
As usual, the terms \emph{redshift} and \emph{blueshift} refer, respectively, to $\zeta_{i \to f} > 0$ and $\zeta_{i \to f} < 0$.\\
Keeping in mind the facts mentioned above, as an example we proceed to determine the redshift factor for a light signal which is emitted by an fundamental observer at a generic point $p_i$ in the plane $\{z \!=\! 0\}$, propagates along a null geodesic $\xi$ of the type analysed in Section \ref{ffSec} (with $E = 0$), and finally reaches an event $p_f$ in the Minkowskian region outside $\To_{\ellp}$ where its frequency is measured by another observer of the same family.
\newpage
\noindent
Firstly, it should be noticed that, in the case under analysis, the frequency at any event $p$ can be expressed as
\begin{equation}
\Omega = \ga\;\Theta(\rho/\!R)\,, \label{omi}
\end{equation}
where $\ga$ is the geodesic momentum defined in Eq. \eqref{gaom}, $\rho$ is the radial coordinate of $p$, and $\Theta$ is a dimensionless function depending on the spacetime parameters $\ellm/R,\ellp/R,a,b$ and on the geodesic coefficient $\om$ (but not on $\ga$). In particular, taking into account Eq.s \eqref{EComp} \eqref{tdfd0} \eqref{tdfd1} and \eqref{freq} \eqref{omi}, one has
\begin{eqnarray}
\Theta(\rho/R) = \mbox{const.} = 1 \,, \qquad \mbox{outside $\To_{\ellp}$}\,, \label{ThetaOut} \vspace{0.1cm} \\
\Theta(\rho/R) = \mbox{const.} = \om/a > 0\,, \qquad \mbox{inside $\To_{\ellm}$}\,. \label{ThetaIn}
\end{eqnarray}
Eq.s \eqref{omi} \eqref{ThetaOut} show that the frequency measured at any event $p_f$ outside $\To_{\ellp}$ is
\begin{equation}
\Omega_f = \mbox{const.} = \ga > 0\,; \label{omf}
\end{equation}
this result, together with Eq.s \eqref{redshift} \eqref{omi}, allows to infer that the redshift factor relative to the said emission $p_i$ and to the measurement $p_f$ reads
\begin{equation}
\zeta_{i \to f} = \Theta(\rho_i/R) - 1\,, \label{zif}
\end{equation}
where $\rho_i$ is the radial coordinate of $p_i$. As a direct consequence of Eq.s \eqref{omi} \eqref{omf}, $\zeta_{i \to f}$ does not depend on the momentum $\ga$ associated to $\xi$. As an example, Fig. 6 shows the plot of the redshift parameter $\zeta_{i \to f}$ as a function of the coordinate $\rho_i/\!R$ corresponding to the emission $p_i$, in the setting of Eq. \eqref{cho1} for $\om = 0.08$.
\\
Consider now a light signal emitted in the Minkowskian region outside $\To_{\ellp}$, performing a time travel along the null geodesic $\xi$, and finally returning outside $\To_{\ellp}$. In this case the previous results \eqref{ThetaOut} and \eqref{zif} yield
\begin{equation}
\zeta_{i \to f} = \mbox{const.} = 0 \qquad \mbox{($p_i$ outside $\To_{\ellp}$)}\,,
\end{equation}
indicating that the light signal is received in the past with no frequency shift with respect to its emission.\\
Conversely, for a light signal emitted inside $\To_{\ellm}$ and observed outside $\To_{\ellp}$, Eq.s \eqref{ThetaIn} and \eqref{zif} imply
\begin{equation}
\zeta_{i \to f} = \mbox{const.} = {\om \over a} - 1 \qquad \mbox{($p_i$ inside $\To_{\ellm}$)}\,. \label{red2}
\end{equation}
The above relation \eqref{red2} makes patent that, for fixed values of the spacetime parameter $a$, either redshift or blueshift can occur, depending on the specific value of the geodesic coefficient $\om$ (which is related, in turn, to the direction of emission in the plane $\{z = 0\}$; see Eq. \eqref{tdfd0}). In this connection, the constraints in Eq. \eqref{impl} (necessary for the considered null geodesic $\xi$ to exist) yield
\begin{equation}
{1 \over b} - 1 \,<\, \zeta_{i \to f} \,<\, {1 \over a}\left(1 - {\ellp \over R} \right) - 1\,. \label{implred}
\end{equation}
A few remarks are in order:
\vspace{-0.15cm}
\begin{enumerate}[(i)]
\item if $b < 1$, then $\zeta_{i \to f} > 0$ for all $\omega$ as in Eq. \eqref{impl}. This means that the light signal under analysis gets unavoidably redshifted;
\vspace{-0.15cm}
\item if $a > 1 \!-\! \ellp/R$, then $\zeta_{i \to f} < 0$ for all $\omega$ as in Eq. \eqref{impl}. In this case, the light signal gets certainly blueshifted;
\vspace{-0.15cm}
\item if $a < 1 \!-\! \ellp/R$ and $b > 1$, then both redshifts and blueshifts are allowed. Actually, for $\om$ as in Eq. \eqref{impl}, the light signal gets redshifted (resp.,\! blueshifted) if $\om < a$ (resp.,\! $\om > a$). The prototypical setting \eqref{cho1} belongs to this case.
\end{enumerate}

\section{The exotic matter content of the spacetime}\label{mattSec}
\vspace{-0.15cm}
Recall that the geometry of the spacetime $\TT$ was established \emph{a priori} in Section \ref{SecGeo}. The corresponding matter content is hereafter deduced \emph{a posteriori} by implementing a ``reversed'' approach to Einstein's equation, originally pursued in \cite{Alcu,Kras} for the construction of exotic spacetimes. In this sense, Einstein's equations (with zero gravitational constant) are deliberately enforced by hand, defining the stress-energy tensor for $\TT$ as the symmetric bilinear form
\begin{equation}
\Te \,:=\, {1 \over 8 \pi G} \left(\Ric\,-\,{1 \over 2}\,\Rs\,g\right), \label{defT}
\end{equation}
where $G$, $\Ric$ and $\Rs$ denote, respectively, the universal gravitational constant, the Ricci tensor and the scalar curvature of the metric $g$. Of course, $\Te$ vanishes identically outside $\To_{\ellp}$ and inside $\To_{\ellm}$, where the spacetime is flat.\\
As usual, the \emph{energy density} measured at any event $p \!\in\! \TT$ by an observer with normalized $4$-velocity $U \in T_p\TT$ (such that $g(U,U) = -1$) is given by
\begin{equation}
\EE(U) \,:=\, \Te(U,U) \,. \label{edens}
\end{equation}
In the following we compute the energy densities measured by two families of observers. Next, we determine the Hawking-Ellis class \cite{Hawk,Viss} of the stress-energy tensor $\Te$.
\vspace{0.15cm}\\
\textbf{The energy density measured by fundamental observers.}
Consider the fundamental observers of Section \ref{SecOr}, with $4$-velocity coinciding with the timelike element $E_{(0)}$ of the tetrad \eqref{SecOr}. In \cite{CQG} it is shown that the energy density measured by such observers at a point $p \in \TT$ of coordinate $(t,\ff,\rho,z)$ can be expressed as
\begin{equation}
\EE\big(E_{(0)}|_p\big) = {1 \over 8 \pi G\,R^2}\; \EFf(\rho/R,z/R) \,, \label{EFf}
\end{equation}
where $\EFf$ is a dimensionless function, depending on the parameters $\ellm/R$, $\ellp/R,a,b$.\\
As an example, consider the setting \eqref{cho1}. Fig. 7 shows the graph of the map $\rho/\!R\! \mapsto\! \EFf(\rho/\!R,0)$. In the same setting it is $\min\!\EFf \!=\! \EFf(1.69...,0) \!\sim\! -2 \cdot 10^4$, yielding
\begin{equation}
\min_{p \in \TT} \EE\big(E_{(0)}|_p\big) \sim -\,{10^{27} \over (R/\me)^2}\; {\gr/\cm^3}\,. \label{EfH2O}
\end{equation}
This makes patent that $\EE(E_{(0)}|_p) \!<\! 0$ for some $p \!\in\! \TT$, which proves that the weak (and, hence, the dominant) energy condition is violated; by similar computations it can be inferred that the strong condition fails as well. Moreover, the numerics in Eq. \eqref{EfH2O} reveal that (the absolute value of) the energy density, while considerably smaller than the Planck density $\rho_P \!:=\! c^5/(\hbar\, G^2)\sim 10^{93} \gr/\cm^3$, is huge on human scales unless $R \gtrsim 10^{13} \me$.
\vspace{0.15cm}\\
\textbf{The energy density measured during a time travel by free fall.}
Let $\xi$ be a timelike geodesic (with tangent $\dot{\xi}$) of the type described in Section \ref{ffSec}. The energy density measured at proper time $\tau$ by an observer in free fall along $\xi$ is given by (see \cite{CQG})
\begin{equation}
\EE\big(\dot{\xi}(\tau)\big) = {\ga^2 \over 8 \pi G\,R^2}\; \EFg\big(\rho(\tau)/R\big)\,, \label{EFg}
\end{equation}
where $\rho(\tau)$ is the radial coordinate associated to $\xi$ and $\EFg$ is a dimensionless function, depending on $\ellm/\!R,\ellp/\!R,a,b$ and on $\ga,\om$.\\
As an example, refer again to the setting \eqref{cho1} and consider a geodesic with $\ga = 10^5$ (see Eq. \eqref{gaNum}). Fig. 5 shows the graph of the function $\rho/R \mapsto \EFg(\rho/R)$ for $\xf = 2 \cdot 10^{-20}$ (see Eq. \eqref{eq1}).
The following numerical results are extracted form Tables 1-3 of \cite{CQG} (cf. also the preceding Eq.s \eqref{eq1}-\eqref{eq3} and \eqref{aq1}-\eqref{aq3}):
\begin{eqnarray}
\min_{\tau \in [0,\tau_2]} \EE\big(\dot{\xi}(\tau)\big) \sim -\,7 \cdot 10^{31}\, \gr/\cm^3 \qquad & \mbox{for $R = 10^2\,\me$\,, $\xf = 10^{-20}$}\,; \vspace{-0.1cm} \\
\min_{\tau \in [0,\tau_2]} \EE\big(\dot{\xi}(\tau)\big) \sim -\,7 \cdot 10^{13}\, \gr/\cm^3 \qquad & \mbox{for $R = 10^{11}\, \me$\,, $\xf = 10^{-11}$}\,; \vspace{-0.25cm} \\
\min_{\tau \in [0,\tau_2]} \EE\big(\dot{\xi}(\tau)\big) \sim -\,0.7\, \gr/\cm^3 \qquad & \mbox{for $R = 10^{18}\, \me$\,, $\xf = 10^{-4}$}\,.
\end{eqnarray}
Comments analogous to those reported below Eq. \eqref{EfH2O} can be made even in the present situation. In particular, the above numerics show that, for $\ga \!=\! 10^5$, the energy densities measured during a time travel by free fall are comparable to those of everyday experience only for enormous values of the radius $R$.
\begin{figure}[t!]\label{fig:EnfEng}
\vspace{-0.7cm}
    \centering
        \begin{subfigure}[b]{0.475\textwidth}\label{fig:Enf}
                \includegraphics[width=\textwidth]{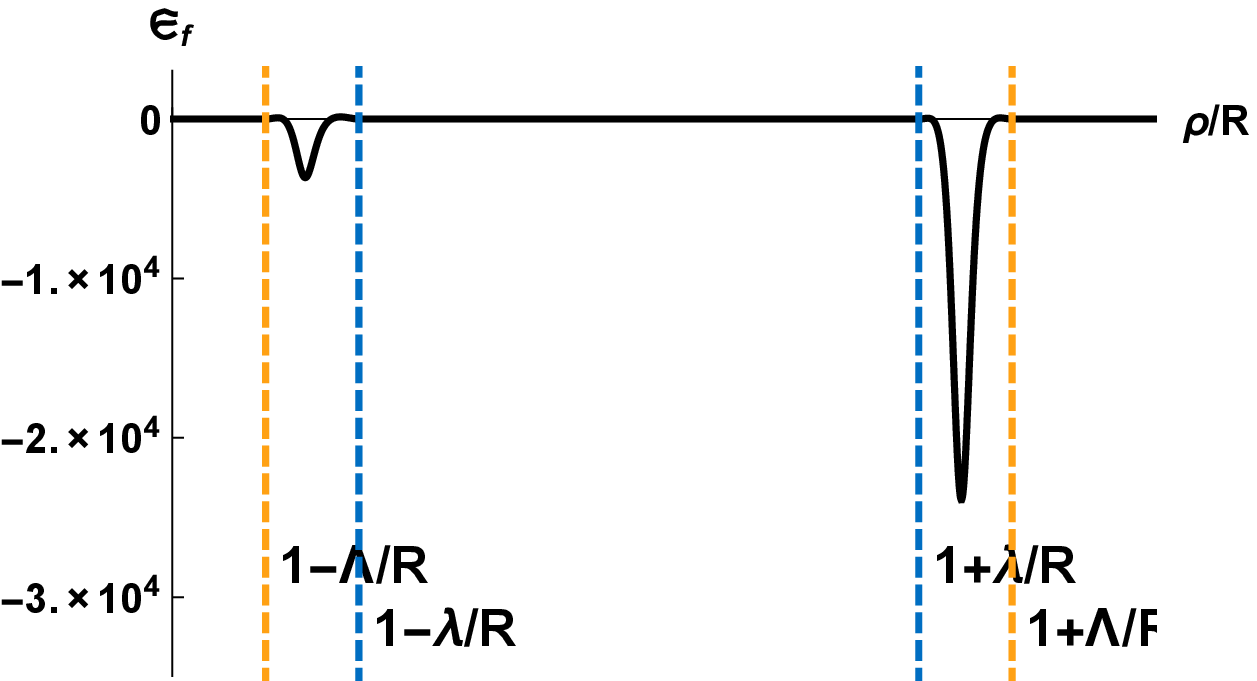}\vspace{-0.cm}
                \caption*{\textsc{Figure 7.} Plot of the function $\EFf(\rho/\!R,0)$ (cf. Eq.s \eqref{cho1} 						\eqref{EFf}).}
        \end{subfigure}
        \hspace{0.4cm}
        \begin{subfigure}[b]{0.475\textwidth}
                \includegraphics[width=\textwidth]{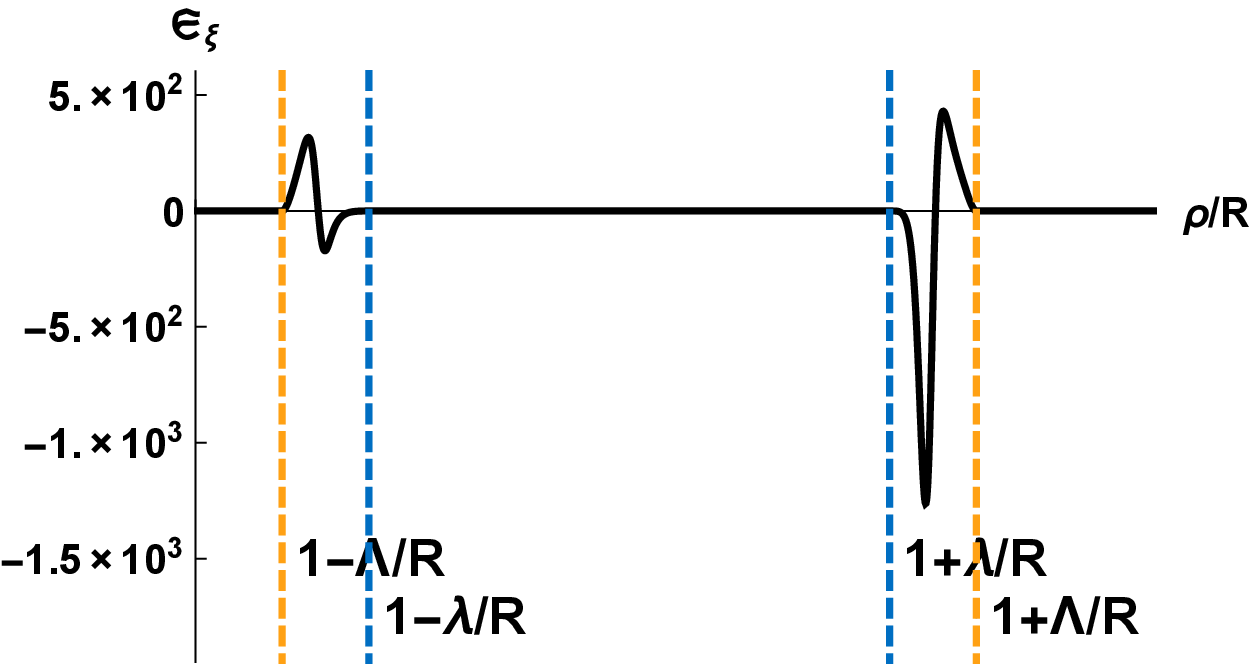}\label{Eng}
                \caption*{\textsc{Figure 8.} Plot of the function $\EFg(\rho/\!R)$ (cf. Eq.s \eqref{cho1} \eqref{EFg}) 					for $\xf = 2 \cdot 10^{-20}$.}
        \end{subfigure}
\end{figure}
\vspace{0.15cm}\\
\textbf{Hawking-Ellis classification of the stress-energy tensor.}
Let $(E_{(\al)})_{\al \in \{0,1,2,3\}}$ be the pseudo-orthonormal tetrad of Section \ref{SecOr} and consider the stress-energy components
\begin{equation}
\Te_{\al\be} := \Te\big(E_{(\al)},E_{(\be)}\big) \qquad \mbox{{\small $(\al,\be \in \{0,1,2,3\})$}}\;. \label{Teab}
\end{equation}
Incidentally, notice that $\Te_{00}$ coincides by definition with the previously discussed energy density $\EE(E_{(0)})$ measured by a fundamental observer.\\
It can be checked by direct inspection that $(\Te_{\al\be})$ is a block matrix of the form
\begin{equation}
(\Te_{\al\be}) = \left(\!
\begin{array}{cc}
\PP		&	\zer	\\
\zer	&	\QQ		\\
\end{array}\!
\right), \qquad \PP = \PP^T, \;\QQ = \QQ^T,
\end{equation}
where $\zer$ is the zero $2 \!\times\! 2$ matrix and $\PP,\QQ$ are $2 \!\times\! 2$ real symmetric matrices.
The (Segre-Pleba\'{n}ski-)Hawking-Ellis classification \cite{Hawk,Viss} relies on the analysis of the eigenvalues $\ka$ determined by the characteristic equation
\begin{equation}
0 = \det\!\big(\Te_{\al\be} - \ka\,\eta_{\al\be}\big) = 
\det(\PP - \ka\,\JJ) \cdot\, \det (\QQ - \ka\,\II)\,,
\end{equation}
where $(\eta_{\al \be}) \!:= \mbox{diag}(-1,1,1,1)$, $\JJ\!:= \mbox{diag}(-1,1)$ and $\II \!:= \mbox{diag}(1,1)$. \\
The study of the $\QQ$-block is elementary. In fact, $\QQ$ is symmetric with respect to the standard Euclidean product on $\R^2$, represented by $\II$; so, by the spectral theorem it follows that the equation $\det(\QQ - \ka\,\II) = 0$ has two (possibly coinciding) real solutions, which correspond to orthogonal spacelike eigenvectors of $(\Te_{\al\be})$.\\
Regarding the $\PP$-block, consider the related discriminant
\begin{equation}
\Delta := (\Te_{00}\!+\!\Te_{11})^2 - 4 \Te_{01}^2\,, \label{Delta}
\end{equation}
where the components $\Te_{00},\Te_{01},\Te_{11}$ are as in Eq. \eqref{Teab}. It can be checked that the eigenvalues determined by $\det(\PP \!-\! \ka\,\JJ) = 0$ are both real if $\Delta \!\geqs\! 0$ (coinciding if $\Delta = 0$) and complex conjugate if $\Delta \!<\! 0$. As a matter of fact, the sign of $\Delta$ varies, depending on the parameters $\ellm/\!R,\ellp/\!R,a,b$, and on the shape function $\XX$. As an example, consider the setting \eqref{cho1}; Fig. 9 shows the plot of $(\mbox{sgn} \Delta)(\rho/\!R,z/\!R)$ and reveals that $\Delta$ can be either positive or negative.
\begin{figure}[t!]\label{fig:HorPot}
\vspace{-0.7cm}
    \centering
                \includegraphics[width=5cm]{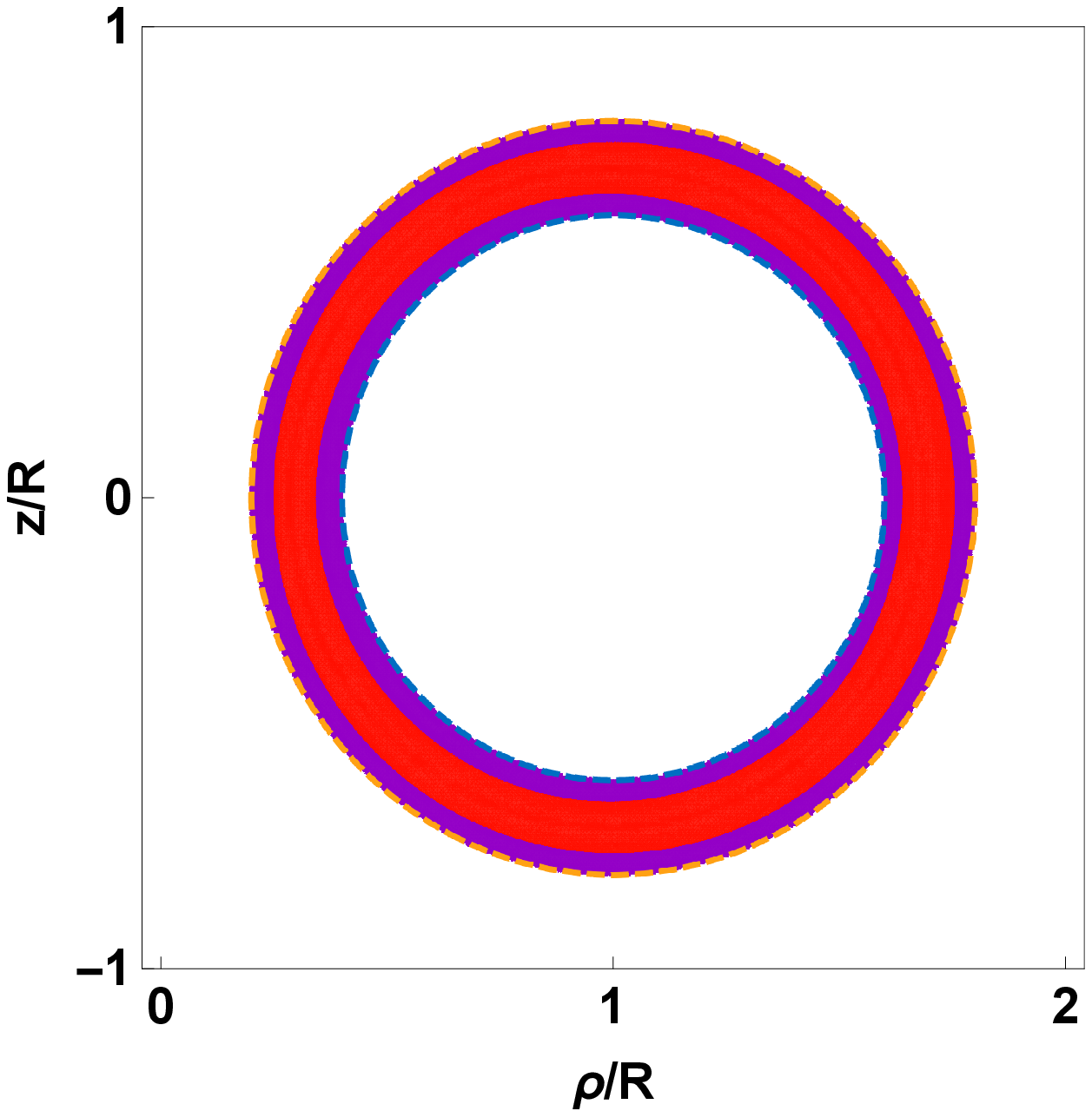}\vspace{-0.1cm}
               	\caption*{\textsc{Figure 9.} Representation of the regions where $\mbox{sgn} \Delta = 1$ (in red) and 
                $\mbox{sgn}\Delta = -1$ (in purple) (cf. Eq.s \eqref{cho1} \eqref{Delta}).}
\vspace{-0.2cm}
\end{figure}
\noindent
More precisely, the equation $\Delta = 0$ implicitly defines two surfaces, yielding a three-layered partition of the curved spacetime region delimited by $\To_{\ellm}$ and $\To_{\ellp}$: 
\vspace{-0.05cm}
\begin{enumerate}[(i)]
\item in the \emph{intermediate layer} it is $\Delta > 0$, so $\det(\PP - \ka\,\JJ) = 0$ has two distinguished real solutions. In view of the previous results for the $\QQ$-block, this means that in this layer the stress-energy tensor $\Te$ fits into \emph{type I} of the Hawking-Ellis classification, namely, the one corresponding to most common matter forms.
\vspace{-0.15cm}
\item in the \emph{inner} and \emph{outer layers} there holds $\Delta < 0$, so $\det(\PP \!-\! \ka\,\JJ) = 0$ has two complex-conjugate solutions. Recalling again the remarks about the $\QQ$-block, this implies that in these layers $\Te$ fits into \emph{type IV} of the Hawking-Ellis classification. 
Stress-energy tensors of this type are known to suffer serious interpretation problems in terms of classical matter sources; nonetheless, the renormalized expectation values of the stress-energy tensor are of this type in a number of semiclassical models for quantum fields on curved backgrounds (see \cite{Viss} and references therein).
\vspace{-0.15cm}
\item at the \emph{transition surfaces} delimiting the previous layers (where $\Delta = 0$), the equation $\det(\PP \!-\! \ka\,\JJ) = 0$ has two coinciding real solutions. Thus, $\Te$ fits into \emph{type II} of the Hawking-Ellis classification, which admits a classical interpretation in terms of radiation.
\end{enumerate}

\section{Future investigations}\label{secfut}
\vspace{-0.2cm}
This conclusive section hints at some possible developments related to the spacetime $\TT$, whose investigation is deferred to future works.
\vspace{0.1cm}\\
\textbf{Accelerated trajectories.} The alleged time travels considered in \cite{CQG} and in Section \ref{ffSec} of this work exclusively refer to observers in free fall along specific timelike geodesics.
Some troublesome features related to such trajectories could be partly overcome by means of accelerated motions. For example, tiny accelerations could allow to avoid the fine tuning issue for the parameter $\xf$ of Eq. \eqref{defxf}, required at the present stage to return to the same spatial position.
Moreover, accelerating observers could travel into the past of significantly larger amounts of time, possibly averting the need to produce enormous (yet, negative) energy densities. To this purpose, their journey should be scheduled as follows: 
(i) freely fall inside $\To_{\ellm}$, following a timelike geodesic like those described in Section \ref{ffSec}; 
(ii) once inside $\To_{\ellm}$, produce a small acceleration so as to end upon a nearby geodesic along which the decrease of the coordinate time $t$ is somehow optimal;
(iii) follow the said geodesic for an arbitrarily large interval of proper time, thus moving into the past with respect to $t$;
(iv) produce again a small acceleration, in order to return on a nearby geodesic connecting the interior of $\To_{\ellm}$ to the Minkowskian region outside $\To_{\ellp}$.\vspace{-0.05cm}
\newpage
\noindent
\textbf{Causal structure and creation of closed causal curves.} The results of Section \ref{ffSec} indicate that there exist (non-necessarily geodetic) closed causal curves self-intersecting at any spacetime point $p\in\TT$. Thus, $\TT$ is an ``eternal time machine'' (in the sense of \cite{Eche,KraB}) where causality is violated at any point. A clearer understanding of the causal structure of $\TT$ could be achieved characterizing its past and future null infinities. In this connection, it would also be of interest to consider a more realistic variant of $\TT$, evolving from an initial non-exotic chronal region; such a variant could be described in rather simple terms assuming the shape function $\XX$ of Eq. \eqref{Xidef} to depend also on the coordinate $t$ and, possibly, on $\ff$ (in addition to $\rho,z$). Conceivably, this alternative model would present a chronology horizon, delimiting the spacetime region with closed causal curves.
\vspace{0.1cm}\\
\textbf{Classical and quantum fields propagating on the background $\TT$.}
A fundamental issue typically affecting spacetimes with CTCs is the stability under back-reaction effects related to (classical and quantum) fields propagating on the said curved backgrounds; namely, such effects are expected to drastically alter the structure of the spacetime under analysis, intrinsically preventing the formation of closed causal curves. This type of arguments led Hawking to formulate of the so-called ``Chronology Protection Conjecture'' \cite{CPC}.
A first step towards the investigation of the stability issue for $\TT$ would be the analysis of the Cauchy problem for the wave equation on it, describing the propagation of a scalar field. In this connection, the construction of the causal propagator comprises non-trivial problems of micro-local analysis, which could be partly eased by the specific symmetries of $\TT$.

\subsection*{Acknowledgment}
\vspace{-0.15cm}
I wish to thank Livio Pizzocchero for valuable comments and suggestions.
This work was supported by: INdAM, Gruppo Nazio\-nale per la Fisica Matematica; ``Progetto Giovani GNFM 2017 - Dinamica quasi classica per il modello di polarone'' fostered by GNFM-INdAM; INFN, Istituto Nazionale di Fisica Nucleare.


\end{document}